# Broadband Magnetic-Manipulated Spintronic Terahertz Emitter with Arbitrarily Tunable Polarizations

Xiaojun Wu[1,6*†], Deyin Kong[1,6†], Tianxiao Nie[2*], Bo Wang[3,5], Meng Xiao[1], Chandan Pandey[2], Yang Gao[1,6], Lianggong Wen[4], Weisheng Zhao[2,4*], Cunjun Ruan[1,6], Jungang Miao[1,6], Li Wang[3], and Yutong Li[3,5]

**Affiliations:**

[1]School of Electronic and Information Engineering, Beihang University, Beijing, 100083, China

[2]Fert Beijing Institute, BDBC, and School of Microelectronics, Beihang University, Beijing 100191, China

[3]Beijing National Laboratory for Condensed Matter Physics, Institute of Physics, Chinese Academy of Sciences, Beijing 100190, China

[4]Beihang-Goertek Joint Microelectronics Institute, Qingdao Research Institute, Beihang University, Qingdao, 266000, China

[5]School of Physical Sciences, University of Chinese Academy of Sciences, Beijing 100049, China

[6]Beijing Key Laboratory for Microwave Sensing and Security Applications, Beihang University, 100191, China

*To whom correspondence should be addressed. E-mail: xiaojunwu@buaa.edu.cn; nietianxiao@buaa.edu.cn; weisheng.zhao@buaa.edu.cn

†Equal contribution.

**Abstract**: Flexible manipulation of terahertz-wave polarization during the generation process is very important for terahertz applications, especially for the next-generation on-chip functional terahertz sources. However, current terahertz emitters could not satisfy such demand, hence calling for new mechanism and conceptually new terahertz source. Here we demonstrate a magnetic-field-controlled, highly-efficient, cost-effective, and broadband terahertz source with flexible switch of terahertz polarization states in ferromagnetic heterostructures driven by femtosecond laser pulses. We verify that the chirality, azimuthal angle, and ellipticity of the generated elliptical terahertz waves can be independently manipulated by delicately engineering of the external applied magnetic fields via effectively manipulating the photo-induced spin currents. Such an ultrafast photomagnetic interaction-based, magnetic-field-controlled, and broadband tunable solid-state terahertz source integrated with terahertz polarization tunability function not only has the capability to reveal physical mechanisms of femtosecond spin dynamics, but also demonstrates the feasibility to realize novel on-chip terahertz functional devices, boosting the potential applications for controlling elementary molecular rotations, phonon vibrations, spin precessions, high-speed terahertz communication, and accelerating the development of ultrafast terahertz opto-spintronics.

## Introduction

Emerged in 1990s, terahertz science and technology benefiting from a variety of intrinsic unique properties of the special frequency range (0.1-30 THz), has experienced an unprecedented rapid progress with potential in fundamental sciences and real applications, especially under the fast development of femtosecond laser technology[1]. Recently, terahertz techniques have also been employed as powerful tools for triggering other novel related researches and applications, such as electron accelerations[2], nonlinear photonics[3], strong-field nonlinear phononics[4], non-ionization imaging[5] and so on. Although this technology has been widely and successfully used in laboratories, the potential of terahertz technology is significant broader, as envisioned in its real applications such as non-destructive spectroscopy[6] and ionization free imaging, highly sensitive sensing[7], short-range wireless communication at terahertz bit rates[8,9] and so on. However, hindering the development of this fascinating field from laboratory researches to real applications lies in the lack of highly efficient terahertz sources, versatile functional devices sensitive detectors, as well as compact and robust systems. Among them, terahertz sources are of significant importance, especially for those integrated with flexible manipulation of terahertz polarization state. As well documented, when carrying with optical spin and angular momentum[10], it enables another freedom of utilizing terahertz waves and boosts the related applications in physics, material sciences, chemistry and biology. However, terahertz sources and functional devices for tuning polarization normally work independently in narrowband. For example, electrically, optically, or temperature-controlled terahertz polarizers and devices that are based on metamaterials[11–14], liquid crystals[15], or segmental quarter-wave plates[16] are always inserted between the source and measured samples to realize their functions. It has turned out that a very smart way is to control the emitted terahertz polarization state before radiation or during the generation process, like two-color laser plasma interaction-based[17–20] or semiconductor-based terahertz sources[21], two laser beams pumped nonlinear crystals[22] or NiO antiferromagnetic terahertz sources[23]. However, laser plasma-based polarization tunable terahertz sources always require high energy laser pumping while suffering relatively low optical-to-terahertz energy conversion efficiency, and semiconductor-based magnetic-field-enhanced terahertz emitters requires more than several Tesla external magnetic fields. Meanwhile, semiconductor (GaAs or InGaAs)-based terahertz antennas[24] suffer the disadvantage of lacking the freedom of high tunability since the terahertz radiation polarization state is imprinted by the designed and fabricated structures. Two laser beams pumped polarization-tunable sources require carefully controlling the phase and amplitude of the two pumping beams. Therefore, there are still plenty spaces to explore polarization-tunable terahertz sources especially integrated with the merits of low-cost, high efficiency and broadband tunability.

Previously, terahertz waves carrying angular momentum properties have been widely used as non-thermal spin injector for investigating spin-galvanic effect[10], photo-galvanic effect[25], and made tremendous contributions to the development of spintronics. Interestingly, recently demonstrated nanofilm terahertz emitters that are based on ferromagnetic heterostructures driven by femtosecond laser pulses have turned out to be a very highly-efficient, cost-effective, ultrabroadband terahertz sources[26]. Under the extremely short ~10 fs laser pulse excitation[27], the ferromagnetic nanofilms can emit terahertz waves with efficiency comparable to the nonlinear crystals (ZnTe, GaP) based emitters, and semiconductor based photoconductive antennas. The ferromagnetic terahertz emitters can not only deliver the aforementioned superior emission characteristics, but also accompany the merit of tunable polarization controlled by the external applied magnetic fields[28].

According to the inverse spin Hall effect (ISHE), the laser injected spin-polarized current in ferromagnetic nanofilms flows to the heavy metals (HMs) leading to the generation of an in-plane charge current[26]. The converted charge currents are expressed by

$$\vec{j}_c^{\perp} = \gamma \vec{j}_s \times \vec{M} / |\vec{M}| \tag{1}$$

where $\vec{j}_c^{\perp}$ is the converted charge current perpendicular to the magnetization; $\gamma$ is the magnitude of spin Hall angles with the sign determined by the type of HMs; $\vec{j}_s$ is the injected spin current density determined by the pump laser fluence; $\vec{M}$ is the magnetization. According to the Maxwell equation, the femtosecond time-varied current emits electromagnetic waves in terahertz frequency range. With this method, experimental results show that the generated terahertz pulses are always linearly polarized with its electric fields perpendicular to the external applied uniform magnetic field direction[29]. Hence rotating the magnetic fields, the emitter also works as an imprinted wire-grid polarizer. However, up to now, it has turned out to be very challengeable to realize arbitrary polarization control and manipulation in such ferromagnetic heterostructures driven by femtosecond laser pulses.

In our work, we intentionally design the external magnetic field distribution and demonstrate broadband elliptical terahertz wave generation in W/CoFeB/Pt trilayer heterostructures. Through tailoring the magnetic fields on purpose, we can efficiently control the chirality as well as the azimuthal angle in the emitted broadband terahertz frequency range. We believe that our demonstrated magnetic controlled broadband terahertz generation not only benefits for deeply understanding the ultrafast spin dynamics but also have values for the next-generation novel terahertz sources and devices.

**Concepts and models**

Inspired by the magnetic control of linearly polarized terahertz generation, it is possible to generate elliptical terahertz beam with controllable chirality, azimuthal angle as well as ellipticity by tailoring the magnetization distributions. Figure 1 shows the concept and theoretical simulated results. For a uniform magnet field distribution, as shown in Figure 1b, linearly polarized terahertz waves can be generated with an electric field perpendicular to the magnetization direction. To generate an elliptically polarized terahertz beam, its two orthogonal electric field amplitudes of $E_x$ and $E_y$ as well as the corresponding phase difference is necessary. When a curved magnetic field goes through uneven regions in the ferromagnetic heterostructures, the dominated magnetic field components at these regions are different, as shown in Figure 1a. When femtosecond laser pulses illuminate onto the left and right uneven regions, there will generate two spin converted in-plane transverse ultrafast charge currents with their predominant flowing directions perpendicular to the corresponding dominated magnetic fields. The phase difference of the two emitted electric field components can be tuned by sample thickness and conductivity, therefore, leading to the generation of elliptical terahertz beam. The detailed theoretical deduction is given in the Supplementary Information. To verify our theory, we first apply our aforementioned model to the uniform magnetic field distribution drawn in Figure 1b, the dominated magnetic fields in different areas are in the same direction. Hence, it can only generate linearly polarized terahertz waves, which agrees very well with previously demonstrated experimental results that the polarization of the terahertz wave is perpendicular to the uniform magnetic field. When we intentionally design nonuniform and twisted magnetic field distribution such as illustrated in Figure 1c, the femtosecond laser pulses induced spin converted photocurrents have the ability to emit elliptically polarized terahertz waves with tuned azimuthal angle, ellipticity and chirality.

Figure 1d-f illustrates the Lissajous curves of the simulated elliptically polarized terahertz waves according to basic theory. Figure 1d exhibits the manipulation of the ellipticity and azimuthal angle of the generated terahertz waves by varying the phase difference from 0 to $\pi$. When the phase difference is $\pi/2$, the generated terahertz wave is circularly polarized, while when it is 0 and $\pi$, the waves are linearly polarized. If the phase changed from $\pi$ to $2\pi$, the $E_y$ component is multiplied by -1, the Lissajous curves is the bottom-up format of the Figure 1d, and the rotation direction is counterclockwise. Thus, the chirality of the emitted waves in our samples can be switched by changing the magnetic field distribution. When we fix the phase difference of $E_x$ and $E_y$ at $\frac{\pi}{3}$, and $E_y = 1$, while varying $E_x$ amplitude, the ellipticity and azimuthal angle of the generated terahertz beam can also be manipulated (see Figure 1e). By tuning the amplitude, we can also control the generation of linearly polarized terahertz waves, as depicted in Figure 1f. Therefore, in principle, we can engineer the external applied magnetic field distribution to generate and manipulate arbitrarily polarization of terahertz waves in ferromagnetic heterostructures.

To verify the aforementioned magnetic tailored spintronic terahertz sources with variable terahertz polarization states, we fabricated ferromagnetic heterostructures of W(1.8)/CoFeB(1.8)/Pt(1.8) trilayers sputtered onto 1 mm thick fused silica substrates by magnetron sputtering method. The numbers in the brackets are the thickness of each layer in nanometer. This sample has already been demonstrated with highly efficient, ultrabroadband and cost-effective terahertz emitter due to ISHE and Fabry-Perot effect in previous work. The elliptical terahertz beam generation experimental schematic diagram is plotted in Figure 2 which are consisted of the W/CoFeB/Pt heterostructure-based terahertz emitter, two-polarizer combined polarization-resolved detection system, an electro-optic sampling (EOS) diagnostic system[30]. The polarization resolved terahertz emission spectroscopy is driven by a commercial Ti:sapphire laser source with 800 nm central wavelength, 100 fs pulse duration, and 80 MHz repetition rate. In our case, the generated terahertz yield is one fourth of that from 1mm thick ZnTe under the similar pumping parameters with 4 µJ/cm², which indicates that the spintronic terahertz emitter has the potential as commercial sources in terahertz spectrometers.

**Flexibly manipulating the chirality and azimuthal angle**

According to our aforementioned model and the theoretical derivation, we first successfully realize the chirality manipulation of the generated terahertz waves in W/CoFeB/Pt by varying the twisted magnetic field distribution. When the magnetic field is bended, we generate elliptically polarized terahertz beams with left-handed (δ⁺) elliptically polarized terahertz waves, while right-handed (δ⁻) elliptically polarized terahertz waves are obtained when changing the curved magnetic fields. Figure 3b and d show the experimentally obtained three-dimensional time-domain terahertz signals with different chirality of left-handed (δ⁺) and right-handed (δ⁻) properties, respectively. The maximum amplitudes of $E_x$ and $E_y$ for the left-handed (δ⁺) polarization are ~3.5, while they are ~3.9 for the right-handed ($\delta^-$). The different ratios of $E_x$ and $E_y$ obtained in our two measurements implies that we have the capability to control the generation of different orthogonal amplitudes by tailoring the magnetic field distribution due to ISHE. Furthermore, we can flexibly manipulate the generation of chiral terahertz beams. Figure 3c shows the Lissajous curve of the two measured terahertz signals at ~1 THz. The ellipticity of the LEP is ~0.021, while REP is ~0.096. The asymmetry of the LEP and REP observed in our experiment is attributed to the uneven sample morphology.

The azimuthal angle of the generated elliptical terahertz waves is very important for specific applications. Therefore, freely controlling the azimuthal angle becomes one of the targets in our experiments. Surprisingly, it can be easily manipulated by rotating the sample together with the magnets, illustrated in Figure 4a and b. Figure 4c summarizes the measured rotation of the azimuthal angles for different elliptical terahertz signals in time-domain. From this figure, we can see that the azimuthal angles can be arbitrarily rotated along with the z-axis with the help of our proposed method. In our model, the major axis of the ellipse is determined by the direction of the combination of external inhomogeneous magnetic fields. Therefore, the rotation direction of the external nonuniform magnetic field can directly manipulate the azimuth angle of the ellipse.

Table 1 shows the absolute value of the emitted terahertz electric field peaks in time-domain. The ratio of the $E_x$ and $E_y$ changes along with the azimuthal angles. From this table, we can see that, by engineering the nonuniform magnetic field distribution, we can easily manipulate the amplitude of the generated $E_x$ and $E_y$. In our experiments, we obtain the minimum ratio of 0.074, when the azimuthal angle is 174.0°. Surprisingly, this number can be scaled up to 9.73 when the angle is rotated to 80.2°. To generate circularly polarized terahertz beam, it is required equal amplitudes of $E_x$ and $E_y$. In our case, it has already been demonstrated that when the azimuthal angles are 132.4°, we can obtain equal amplitudes of $E_x$ and $E_y$. However, the strictly required phase difference is not that easy to be achieved by controlling the material thickness and conductivity in this ferromagnetic/nonferromagnetic nanofilms, which has to explored in the future.

**Freely tuning of the ellipticity**

To control the ellipticity, one method is to vary the phase difference, as shown in Figure 5b. The other is to change the amplitude by varying the emitting areas, (see Figure 5c). Figure 5a shows the typical sample status with nonuniform magnetic field distribution. Figure 5d shows the ellipticity spectra of the detected terahertz waves with five polarization states. The equation to calculate the ellipticity[17] is following

$$\delta = \pm \frac{b}{a} \quad (2)$$

where $\delta$ is the ellipticity, $a$ is the major axis and $b$ is the minor axis. The ellipticity is positive for the left-handed waves, while negative for the right-handed waves. The linearly polarized signal in Figure 5d are also set to negative because most of the phase difference shown in Figure

5f is approximately a small right-handed elliptically polarized light. The absolute value of ellipticities increase along with the frequency.

Figure 5e illustrates the Fourier transformed spectra and their corresponding phases of five polarization states for different magnetic field distributions. The emission frequency range is from 0.2-2.8 THz, while the maximum emission occurred at the frequency of ~1.06 THz for all the five cases. Here we focus on the phases of $E_x$ and $E_y$, which results in the formation of elliptical terahertz waves. Figure 5f is the phase difference ($\Delta\varphi$) of $E_x$ and $E_y$. For linear polarization ($\delta°$), the $\Delta\varphi$ is near $-\pi$. The error is ~0.1 from 0.4 THz to 2.3 THz. For left-handed elliptical polarization ($\delta^+$ and $\delta^{+\prime}$), the $\Delta\varphi$ is between $-\pi$ and 0. For right-handed elliptical polarization ($\delta^-$ and $\delta^{-\prime}$), the $\Delta\varphi$ is between 0 and $\pi$ ($\delta^-$) or between $-2\pi$ and $-\pi$ ($\delta^{-\prime}$). The errors in $\delta^{-\prime}$ case above 1.3 THz is possibly caused by the extreme low signal-to-noise ratio of $E_y$, shown in the bottom of Figure 5e. All the cases for phase differences start at 0 or $-\pi$ in Figure 5f. They represent the linear polarization in low frequencies. Along with increasing of the frequency, the phase difference varies linearly, indicating higher ellipticity (Figure 5d) for higher frequencies. The possible reason is attributed to the Fabry-Perot effect, and the detail analysis is shown in Supplementary Information. Our experimental results agree very well with our proposed model.

Figure 6 shows the generated polarized terahertz beams in five selected typical frequencies of 0.5, 1.0, 1.5, 2.0 and 2.5 THz for the aforementioned five polarization states. It indicates that the magnetic controlled ferromagnetic terahertz source is ellipticity tunable in a very broadband frequency range which is illustrated in the right insert of Figure 6. With this illustration, it can be further confirmed that the ellipticity of the generated terahertz waves is bigger in the high frequencies than lower frequencies. Actually, in the whole emission frequency range of 0.2-2.8 THz, all the generated terahertz frequencies are deposited with ellipticity properties predicting specific applications with the broadband tunability property. We here demonstrate broadband tunable elliptical terahertz beam generation in ferromagnetic nanofilms by intentionally varying the magnetic field distribution. However, the nanofilm thickness dependent optical-to-terahertz energy conversion efficiency limits the further increase of the sample thickness which makes it challengeable to obtain phase difference of $\pi/2$ for circularly polarized terahertz generation. Deep investigations to produce perfect circularly polarized terahertz waves need to be implemented in the future but we here propose a possible and effective method for the prospective exploration.

**Conclusion and outlook**

We demonstrate, to the best of our knowledge, the first broadband polarization tunable terahertz sources based on a ferromagnetic heterostructures pumped by femtosecond laser pulses. This terahertz source can radiate highly efficient terahertz waves even driven by nJ-level femtosecond laser pulses with 100 fs pulse durations. Furthermore, we can switch the chirality, azimuthal angle, and ellipticity of the generated elliptical terahertz wave through dedicatedly tailoring the applied nonuniform magnetic field distribution. This kind of terahertz sources integrated with polarization tunability function holds various advantages of low cost, compactness, highly efficient, broadband and continuous magnetic control capability, thereby providing significantly more flexibility and overwhelming superiority than previous demonstrated terahertz sources. All these demonstrations are strongly benefiting from the highly efficient inverse spin-orbital interaction in the optomagnetic process[31], making it well-suited for on-chip ultrafast terahertz opto-spintornic devices. Besides, our methods also provide inspirations for the development of various on-chip high-performance terahertz photonic or spintronic devices for complex terahertz spin, terahertz angular momentum, and controlled wavefront integrated terahertz emitters[32,33]. We believe that this work may open an avenue towards the development of various on-chip terahertz opto-spintronic apparatuses based on magnetic materials with nanometer thickness, which will play important roles in future fundamental sciences and real applications.

## Methods

**Sample preparation.** The sample growth was conducted in a high-vacuum AJA sputtering system with a base pressure of $10^{-9}$ Torr. Before the deposition, the substrate was cleaned by Ar plasmonic etching for a dust-free surface. The deposition conditions (i.e. Ar pressure and applied power) were carefully optimized to promise the best quality and reproducibility. The growth rate for W, CoFeB and Pt was 0.21 Å/s, 0.06 Å/s and 0.77 Å/s, respectively. During the deposition, a sample rotation was performed to ensure a good uniformity. Each of the capping layers of W and Pt and the ferromagnetic layer of CoFeB are all 1.8 nm in our case, thus the total thickness of the trilayer is 5.4 nm. The samples are mounted on a custom designed rotator with variable external static weak magnetic fields (~50 mT). In our measurements, we design and perform non-uniform magnetic fields to manipulate the injected spin current distribution in order to control the far-field radiated terahertz polarizations.

**Polarization resolved terahertz time-domain emission spectroscopy.** The femtosecond laser pulses are divided into two beams. One beam with 90% energy of ~400 mW (4 μJ/cm² pump fluence after a mechanical chopper) is used for terahertz generation, while the other is used as a probing beam in a 1 mm thick ZnTe detector for EOS which includes a quarter-wave plate, a Wollaston prism, a balanced detector. The linearly polarized pumping beam normally incidents onto the ferromagnetic heterostructures with a plano-convex lens (focal length=100 mm) to inject spin-polarized currents with the flowing direction along the propagation of the light in z axis. These currents convert into in-plane charge currents when they flow to the heavy metal layer with the assistance of an applied magnetic field, and radiate terahertz pulses with different polarization states. An automatically controlled terahertz polarizer with fixed positive and negative rotation angle of 45 degree with respect to the vertical direction of the laboratory coordinate is employed to resolve the generated terahertz electric field components of $E_1$ and $E_2$ with the help of EOS. The extinction ratio for this polarizer is ~1000. The obtained terahertz electric fields parallel or perpendicular to the magnetic fields are $E_x$ and $E_y$, which are calculated from $E_2 + E_1$ and $E_2 - E_1$, respectively. In order to further verify the sensitivity of the polarization resolved system, we use a ZnTe as an emitter and another terahertz polarizer after the ZnTe emitter to produce pure linearly polarized terahertz pulses. These pulses go through the rotation polarizer and we can only achieve linearly polarized terahertz signal. The negligible $E_y$ in this case is used to confirm the accuracy of the polarization dependent diagnostic method. The optical path from the terahertz emitter to the ZnTe detector are sealed in a vacuum chamber in order to rule out the influence of the water vapor, and all these measurements are performed at room temperature.

**Acknowledgements:** This work is supported by the National Nature Science Foundation of China (Grants No. 11827807, 11520101003, 11861121001, 61831001 and 61775233), the Strategic Priority Research Program of the Chinese Academy of Sciences (Grant No. XDB16010200 and XDB07030300), the National Basic Research Program of China (Grant No. 2014CB339800), the International Collaboration Project (Grant No. B16001), and the National Key Technology Program of China (Grant No. 2017ZX01032101). Dr. Xiaojun Wu thanks the "Zhuoyue" Program and "Qingba" Program of Beihang University (Grant No. ZG216S1807, ZG226S1832, and KG12052501). Dr. Tianxiao Nie thanks the support from the 1000-Young talent program of China.

**Author contributions:** X.J.W., and T.X.N. conceived and coordinate the femtosecond control of spin-charge current conversion and terahertz emission project. D.Y.K., B.W., and Y.G. performed the measurements and analyzed the samples with help from X.J.W. D.Y.K. analyzed the data and draw the figures with help from X.J.W. T.X.N. and C.P. designed and fabricated the samples. The theoretical formalisms were derived by D.Y.K. with help from X.J.W., and T.X.N. With contributions from C.J.R., W.S.Z., J.G.M., Y.T.L., and L.W., X.J.W. and D.Y.K. wrote the paper. All authors discussed the results and commented on the manuscript.

**Figures**

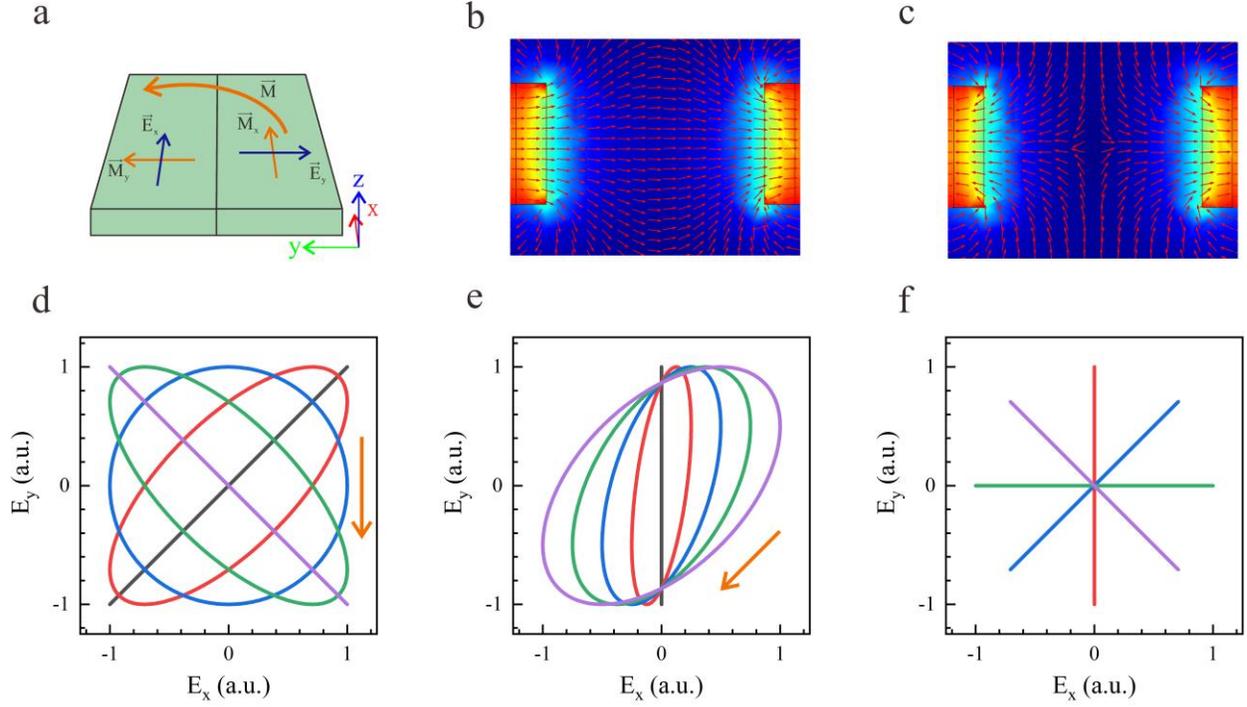

**Figure 1 | Theory derivation of the elliptical terahertz beam generation in ferromagnetic heterostructures via tailoring the magnetic field distribution. a,** Schematic diagram of the twisted magnetic field distribution in two areas of the ferromagnetic heterostructures. x, y, and z are the axes in laboratory coordinate. $M$, $M_x$ and $M_y$ are the magnetic vectors, and their components in x and y directions, respectively. $E_x$, and $E_y$ are the generated terahertz electric field components perpendicular with respect to $M_y$ and $M_x$, respectively. **b and c,** The simulated uniform and non-uniform static magnetic field distributions generated by magnets, respectively. **d,** The Lissajous curves of simulated terahertz polarization when fixing the amplitudes of $E_x = E_y = 1$, while the phase differences ($\Delta\varphi$) are 0 (black), $\frac{\pi}{4}$ (red), $\frac{\pi}{2}$ (blue), $\frac{3\pi}{4}$ (green), and $\pi$ (purple). **e,** The Lissajous curves with fixed $\Delta\varphi$ of $\frac{\pi}{3}$, and $E_y = 1$, while varying $E_x$ amplitude with 0 (black), 0.25 (red), 0.5 (blue), 0.75 (green), and 1.0 (purple), respectively. **f,** Generated linearly polarized terahertz waves with $\Delta\varphi$ of 0, while varying $E_x$ and $E_y$ amplitudes. Blue: $E_x = \sqrt{2}/2$, $E_y = \sqrt{2}/2$; red: $E_x = 0$, $E_y = 1$; purple: $E_x = \sqrt{2}/2$, $E_y = -\sqrt{2}/2$; green: $E_x = 1$, $E_y = 0$.

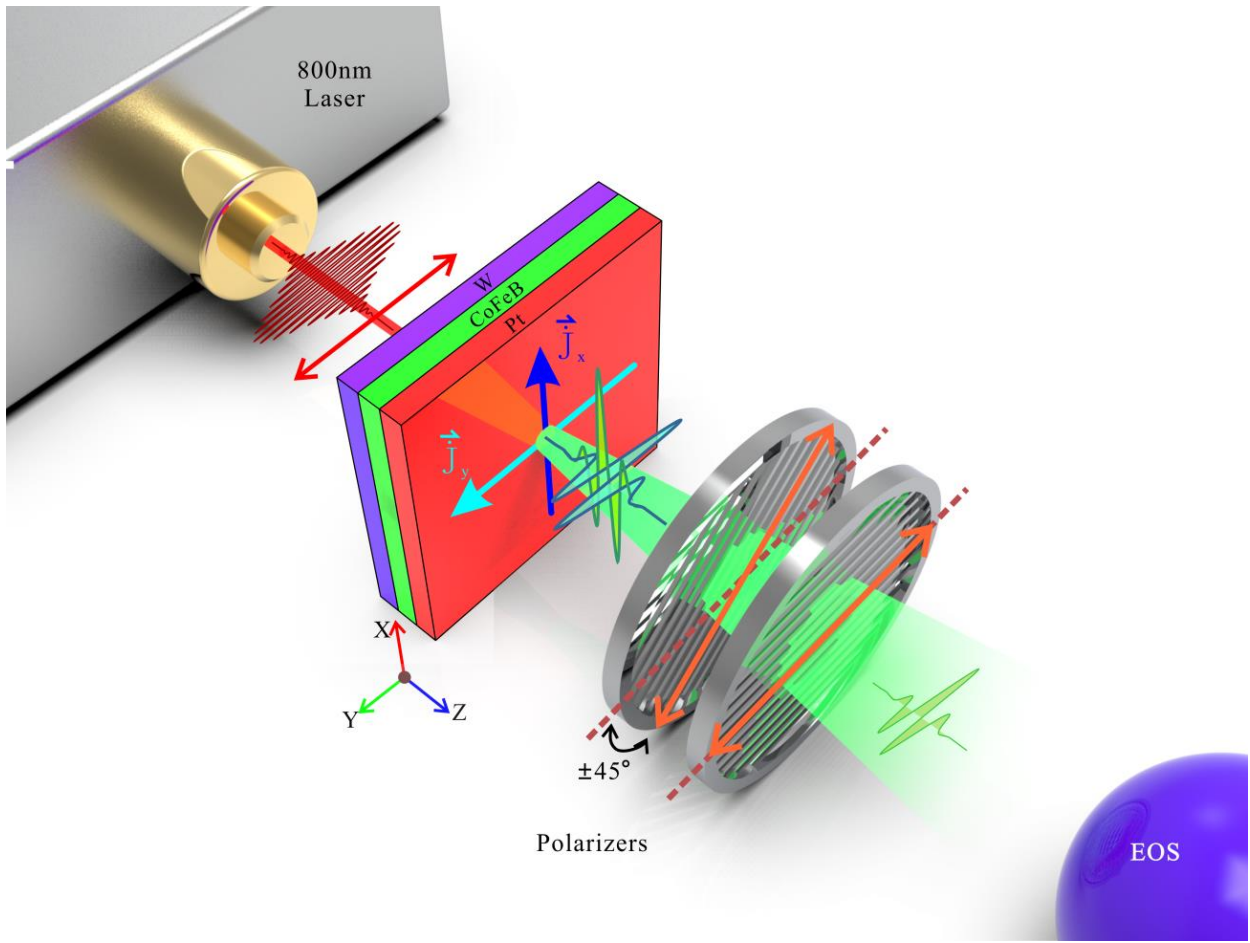

**Figure 2 | Experimental schematic diagram of magnetic manipulated elliptically polarized terahertz beam generation**. Linearly polarized femtosecond laser pulses illuminate onto W/CoFeB/Pt heterostructure nanofilm. The generated photocurrents in x and y directions are $\vec{J}_x$ and $\vec{J}_y$, respectively. The emitted terahertz signal propagates through two terahertz polarizers and diagnosed by polarization resolved EOS.

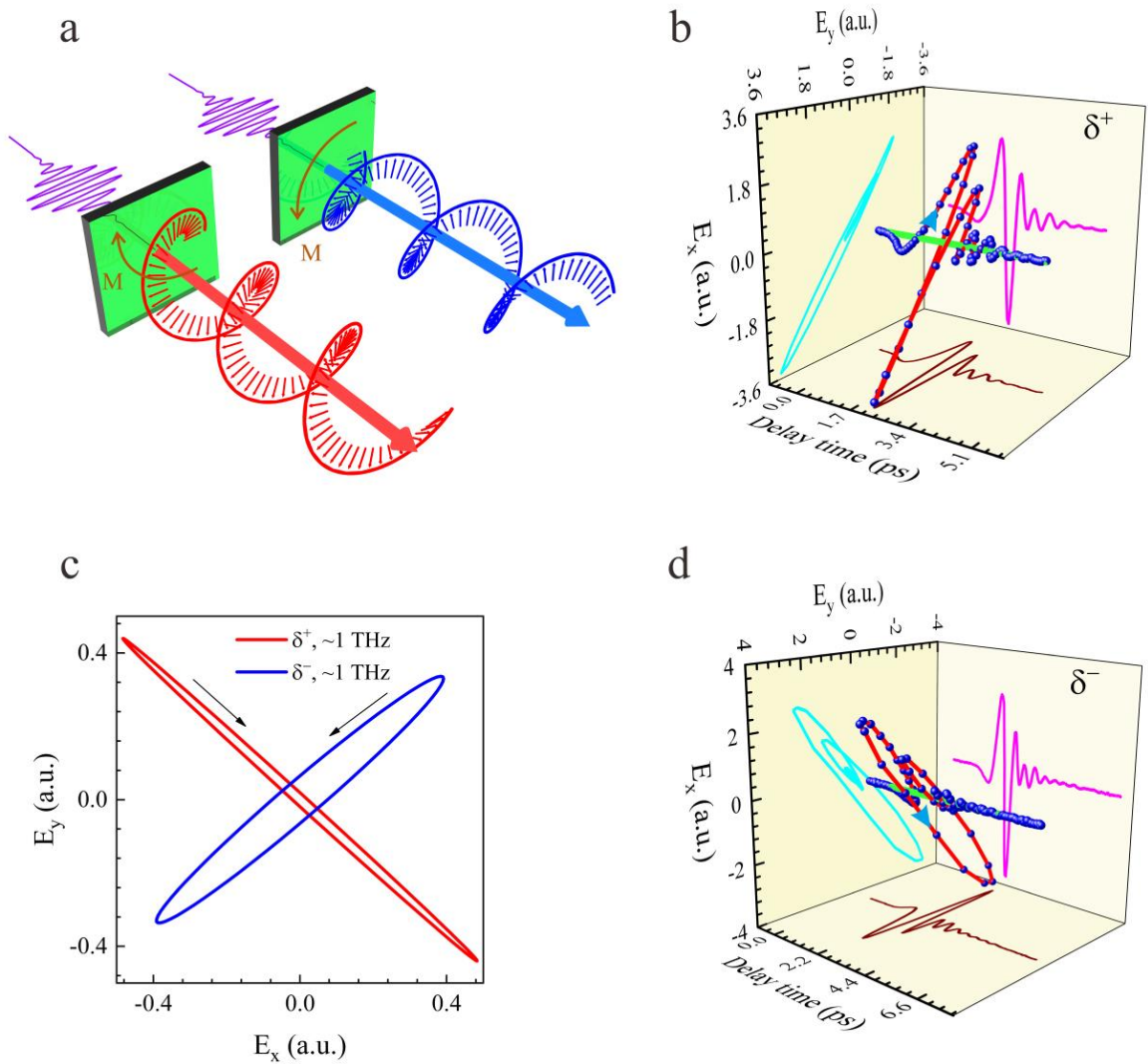

**Figure 3 | Manipulation of the terahertz chirality by changing the twisted magnetic field distribution. a,** Carton of the concept. **b and d,** The measured three-dimensional terahertz temporal waveforms with left-handed and right-handed elliptical polarization states, assumed that the z-axis is in the same direction of the delay time. **c,** The Lissajous curves of the above two signals at ~1 THz.

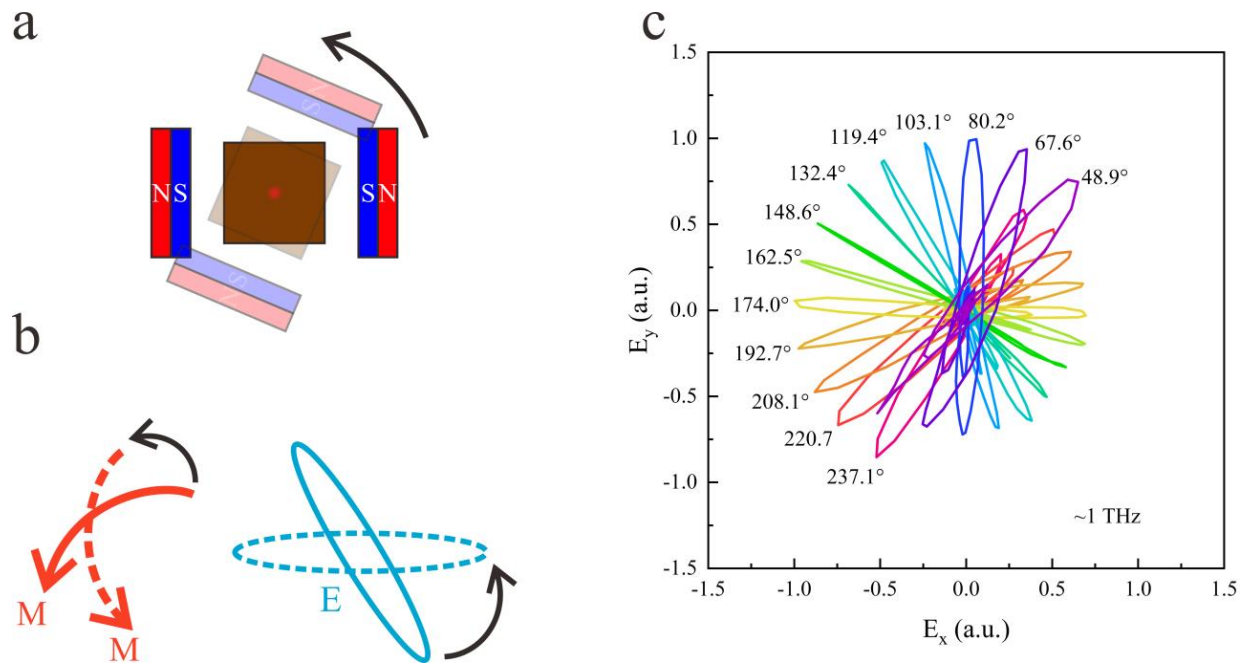

**Figure 4 | Arbitrarily rotating the azimuthal angle of the elliptical terahertz beam. a,** Schematic diagram of rotating sample plus magnetic field. **b,** The major axis of the elliptical beam changes along with the rotation of the magnetic field together with the sample. **c,** The rotation of the terahertz azimuthal. The sample is rotated from 0° to 180°, and the interval is 15°. The amplitudes of the $E_x$ and $E_y$ are normalized. All the signals are left-handed elliptically polarized waves.

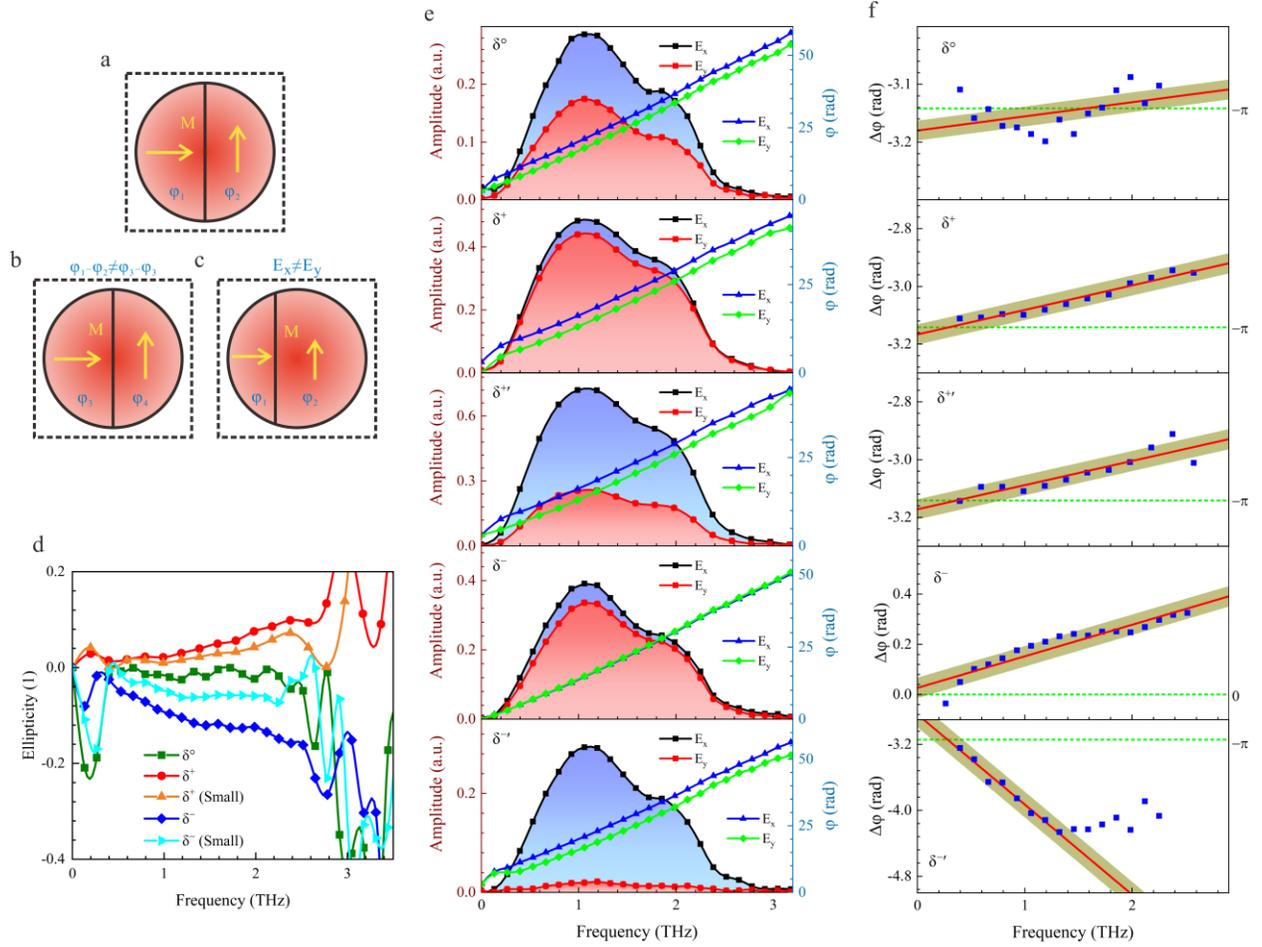

**Figure 5 | The variation of the terahertz ellipticity. a,** A typical status for emitting an elliptically polarized terahertz wave. The phase of the wave from the left ($\varphi_1$) is different from the right ($\varphi_2$). **b,** The variation of the phase difference induced ellipticity change, $\varphi_1 - \varphi_2 \neq \varphi_3 - \varphi_4$. **c,** The change of the emission area induced different ellipticities. **d,** Five terahertz waves with different ellipticities under different magnetic field distribution. **e,** The Fourier transformed spectra and their corresponding phases of the five polarizations. The phases are in the factor of +1 (Science). **f,** The phase difference, $\Delta\varphi = \varphi_y - \varphi_x$, where $\varphi_x$ and $\varphi_y$ are the phases for $E_x$ and $E_y$, respectively. The red curves are the linearly fitting. $\delta^\circ$, $\delta^+$ and $\delta^-$ mean the linearly polarized, the left-handed and right-handed elliptically polarized waves, respectively. $\delta^{+\prime}$ and $\delta^{-\prime}$ are the left-handed and right-handed elliptical terahertz beams with small ellipticities.

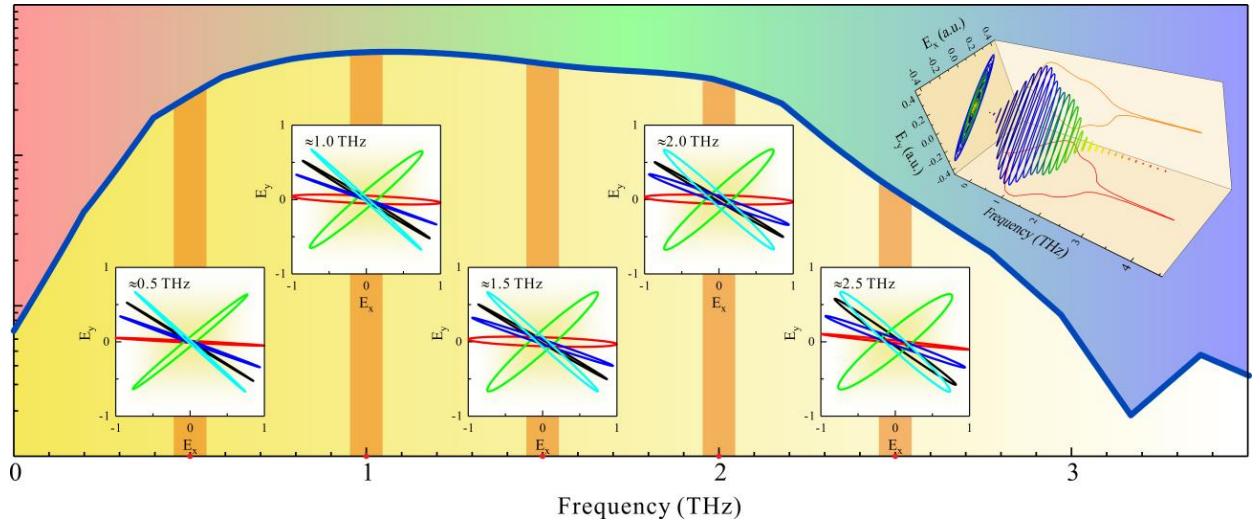

**Figure 6 | Broadband elliptical terahertz beam generation.** The polarization states of the five terahertz waves in Figure 5d at 0.5 THz, 1.0 THz, 1.5 THz, 2.0 THz and 2.5 THz, respectively. Black: $\delta^\circ$; Red: $\delta^{-\prime}$; Green: $\delta^-$; Blue: $\delta^{+\prime}$; Light blue: $\delta^+$. The ellipses are extracted from the Fourier transformed results of the time-domain signals, and the amplitudes are normalized. **Inset:** The three-dimensional broadband polarization spectrum of a typical right-handed elliptically polarized terahertz wave.

**Table**

Table 1 | The peaks of temporal signals of emitted electric field.

| $\theta$ (°) | $|E_{x,\max}|$ (a.u.) | $|E_{y,\max}|$ (a.u.) | $|E_{y,\max}/E_{x,\max}|$ (1) |
|---|---|---|---|
| 48.9 | 0.894 | 1.04 | 1.17 |
| 67.6 | 0.838 | 2.23 | 2.66 |
| 80.2 | 0.430 | 4.18 | 9.73 |
| 103.1 | 1.06 | 4.27 | 4.05 |
| 119.4 | 1.23 | 2.18 | 1.77 |
| 132.4 | 3.17 | 3.38 | 1.07 |
| 148.6 | 2.88 | 1.68 | 0.585 |
| 162.5 | 4.47 | 1.33 | 2.98 |
| 174.0 | 2.84 | 0.208 | 0.074 |
| 192.7 | 2.36 | 0.536 | 0.227 |
| 208.1 | 4.02 | 2.17 | 0.539 |
| 220.7 | 3.70 | 3.32 | 0.896 |
| 237.1 | 2.00 | 3.27 | 1.64 |

# References:

1. Zhang, X. C., Shkurinov, A. & Zhang, Y. Extreme terahertz science. *Nat. Photonics* **11,** 16–18 (2017).
2. Zhang, D. *et al.* Segmented terahertz electron accelerator and manipulator (STEAM). *Nat. Photonics* **12,** 336–342 (2018).
3. Von Hoegen, A., Mankowsky, R., Fechner, M., Först, M. & Cavalleri, A. Probing the interatomic potential of solids with strong-field nonlinear phononics. *Nature* **555,** 79–82 (2018).
4. Hafez, H. A. *et al.* Extremely efficient terahertz high-harmonic generation in graphene by hot Dirac fermions. *Nature* (2018). doi:10.1038/s41586-018-0508-1
5. Mittleman, D. M. Twenty years of terahertz imaging [Invited]. *Opt. Express* **26,** 9417–9431 (2018).
6. Jin, Z. *et al.* Accessing the fundamentals of magnetotransport in metals with terahertz probes. *Nat. Phys.* **11,** 761–766 (2015).
7. Wu, X. *et al.* Alkanethiol-functionalized terahertz metamaterial as label-free, highly-sensitive and specific biosensor. *Biosens. Bioelectron.* **42,** 626–631 (2013).
8. Ma, J., Karl, N. J., Bretin, S., Ducournau, G. & Mittleman, D. M. Frequency-division multiplexer and demultiplexer for terahertz wireless links. *Nat. Commun.* **8,** 729 (2017).
9. Piesiewicz, R. *et al.* Short-range ultra-broadband terahertz communications: Concepts and perspectives. *IEEE Antennas Propag. Mag.* **49,** 24–39 (2007).
10. Ganichev, S. D. *et al.* Spin-galvanic effect. *Nature* **417,** 153–156 (2002).
11. Grady, N. K. *et al.* Terahertz metamaterials for linear polarization conversion and anomalous refraction. *Science.* **340,** 1304–1307 (2013).
12. Gu, J. *et al.* Active control of electromagnetically induced transparency analogue in terahertz metamaterials. *Nat. Commun.* **3,** 1151 (2012).
13. Savinov, V., Fedotov, V. A., Anlage, S. M., De Groot, P. A. J. & Zheludev, N. I. Modulating Sub-THz radiation with current in superconducting metamaterial. *Phys. Rev. Lett.* **109,** 243904 (2012).
14. Kan, T. *et al.* Enantiomeric switching of chiral metamaterial for terahertz polarization modulation employing vertically deformable MEMS spirals. *Nat. Commun.* **6,** 8422 (2015).
15. Wang, L. *et al.* Broadband tunable liquid crystal terahertz waveplates driven with porous graphene electrodes. *Light Sci. Appl.* **4,** e253 (2015).
16. Masson, J.-B. & Gallot, G. Terahertz achromatic quarter-wave plate. *Opt. Lett.* **31,** 265–267 (2006).
17. Lu, X. & Zhang, X. C. Generation of elliptically polarized terahertz waves from laser-induced plasma with double helix electrodes. *Phys. Rev. Lett.* **108,** 123903 (2012).
18. Wang, W. M., Gibbon, P., Sheng, Z. M. & Li, Y. T. Tunable circularly polarized terahertz radiation from magnetized gas plasma. *Phys. Rev. Lett.* **114,** 253901 (2015).

19. Zhang, Z. *et al.* Manipulation of polarizations for broadband terahertz waves emitted from laser plasma filaments. *Nat. Photonics* **12,** 554–559 (2018).

20. You, Y. S., Oh, T. Il & Kim, K.-Y. Mechanism of elliptically polarized terahertz generation in two-color laser filamentation. *Opt. Lett.* **38,** 1034–1036 (2013).

21. Sarukura, N., Ohtake, H., Izumida, S. & Liu, Z. High average-power THz radiation from femtosecond laser-irradiated InAs in a magnetic field and its elliptical polarization characteristics. *J. Appl. Phys.* **84,** 654–656 (1998).

22. Amer, N., Hurlbut, W. C., Norton, B. J., Lee, Y. S. & Norris, T. B. Generation of terahertz pulses with arbitrary elliptical polarization. *Appl. Phys. Lett.* **87,** 221111 (2005).

23. Kanda, N. *et al.* The vectorial control of magnetization by light. *Nat. Commun.* **2,** 362 (2011).

24. Wang, D., Gu, Y., Gong, Y., Qiu, C.-W. & Hong, M. An ultrathin terahertz quarter-wave plate using planar babinet-inverted metasurface. *Opt. Express* **23,** 11114–11122 (2015).

25. Ganichev, S. D. *et al.* Conversion of spin into directed electric current in quantum wells. *Phys. Rev. Lett.* **86,** 4358–4361 (2001).

26. Kampfrath, T. *et al.* Terahertz spin current pulses controlled by magnetic heterostructures. *Nat. Nanotechnol.* **8,** 256–260 (2013).

27. Seifert, T. *et al.* Efficient metallic spintronic emitters of ultrabroadband terahertz radiation. *Nat. Photonics* **10,** 483–488 (2016).

28. Wu, Y. *et al.* High-Performance THz Emitters Based on Ferromagnetic/Nonmagnetic Heterostructures. *Adv. Mater.* **29,** 1603031 (2016).

29. Seifert, T. S. *et al.* Femtosecond formation dynamics of the spin Seebeck effect revealed by terahertz spectroscopy. *Nat. Commun.* **9,** 2899 (2018).

30. Wu, Q. & Zhang, X.-C. Free-space electro-optics sampling of mid-infrared pulses. *Appl. Phys. Lett.* **71,** 1285–1286 (1997).

31. Huisman, T. J. *et al.* Femtosecond control of electric currents in metallic ferromagnetic heterostructures. *Nat. Nanotechnol.* **11,** 455–458 (2016).

32. Zhou, C. *et al.* Broadband Terahertz Generation via the Interface Inverse Rashba-Edelstein Effect. *Phys. Rev. Lett.* **121,** 086801 (2018).

33. Jungfleisch, M. B. *et al.* Control of Terahertz Emission by Ultrafast Spin-Charge Current Conversion at Rashba Interfaces. *Phys. Rev. Lett.* **120,** 207207 (2018).